\documentstyle[12pt]{article}

\newcommand{\Al}{\mbox{\boldmath $\alpha$}}
\newcommand{\ba}{\begin{array}}
\newcommand{\be}{\begin{equation}}

\newcommand{\bea}{\begin{eqnarray}}
\newcommand{\non}{\nonumber}

\newcommand{\ea}{\end{array}}
\newcommand{\ee}{\end{equation}}
\newcommand{\eea}{\end{eqnarray}}

\newcommand{\lm}{\lambda}
\newcommand{\Lm}{\Lambda}

\newcommand{\epl}{\epsilon}

\newcommand{\rf}[1]{(\ref{eq:#1})}

\newtheorem{th}{Theorem}[section]
\newtheorem{prop}[th]{Proposition}
\newtheorem{df}[th]{Definition}

\newtheorem{conj}[th]{Conjecture}

\begin{document}
{\bf
INVARIANTS OF COLORED LINKS AND A PROPERTY OF \\
THE CLEBSCH-GORDAN COEFFICIENTS  OF $U_q(g)$\footnote{
to appear in the Proceedings of the 21st International Conference
on the Differential Geometry Methods in Theoretical Physics,
5-9 June, 1992, Tianjin, China}}
{\vskip 14pt}
\begin{center}
{TETSUO DEGUCHI
and TOMOTADA OHTSUKI $^{\dagger}$}
\end{center}
\date{}
\begin{center}
\it
Department of Physics, Faculty of  Science,\\
     University of Tokyo, Hongo, Bunkyo-ku, Tokyo 113, Japan
\end{center}
\begin{center}
\it
$^{\dagger}$ Department of Mathematical Sciences, \\
     University of Tokyo, Hongo, Bunkyo-ku, Tokyo 113, Japan
\end{center}
\par \noindent
{\bf Abstract}.
\par
We show that multivariable colored link invariants are derived from
the roots of unity representations of $U_q(g)$.
We propose a property of the Clebsch-Gordan coefficients of
$U_q(g)$, which is important for defining the invariants of colored links.
For $U_q(sl_2)$  we explicitly prove the property,
and then construct invariants of colored links and colored ribbon graphs,
which generalize the multivariable Alexander polynomial.

\section{Introduction}

Recently a new family of invariants of colored oriented links
and colored oriented ribbon graphs is introduced, which gives
generalizations of the multivariable Alexander polynomial.$^{1,2,6}$
The new invariants are related to the
 roots of unity  representations of $U_q(sl_2)$
 where $(X^{\pm})^N=0$ and $(K)^N=q^{2Np}$ ($p \in {\bf C}$)
and $q=\epl$.$^{5,6,7}$
Here  $\epl=\exp({\pi i s}/N)$, and the integers $N$ and $s$ are coprime.
 We call the representations nilpotent representations.

The invariants of colored links have a property that
they vanish for disconnected links.$^{2,6}$ Due to this property
 a proper regularization method is necessary
 for definition of the invariants.

In this paper we show an important  property
of the Clebsch-Gordan coefficients (CGC) of the
nilpotent  representations of $U_q(sl_2)$,
which leads to the definition of the colored link invariants.
We consider the nilpotent reps of $U_q(g)$ such that
$(X_i^{\pm})^N=0$, $(K_i)^N=q^{2Np_i}$ ($p_i \in {\bf C}$ ) and $q=\epl$.
We give a conjecture that the property of CGC also holds for
the nilpotent reps of $U_q(g)$ and we can define
invariants of colored links  for $U_q(g)$
in the same way as  $U_q(sl_2)$.

\section{CGC of the nilpotent representations}

We introduce some symbols for a positive integer $n$ and
a complex parameter $p$.
\be
[n]_q  = {\frac {q^{n} - q^{-n}} {q - q^{-1}} } , \quad
[n]_q ! = \prod_{k=1}^{n} [k]_q , \quad
[p;n]_q ! = \prod_{k=0}^{n-1} [p-k]_q .
\ee
For $n=0$ we assume $[0]_q!  =  [p;0]_q! = 1 $.


We introduce  the nilpotent rep. $V^{(p)}$ $^{5}$:
$ \pi(X^{+})^a_{b}=$
${\sqrt{[2p - a]_{\epl} [a+1]_{\epl}}}$
$ \delta_{a+1, b}$,
$ \pi(X^{-})^a_{b} =$
$ {\sqrt{[2p-a+1]_{\epl} [a]_{\epl}}}$ $\delta_{a-1, b}$,
$ \pi(K)^a_{b}$ = $\epl^{(2p - 2a)} \delta_{a, b}$.
Here  $a,b =0,1, \cdots N-1$.
We note that $N$ is related to the dimensions of the representation.
When $N$ is even, $\epl$ is $2N$-th primitive root of unity,
while when $N$ is odd, $\epl$ may be $2N$-th or
$N$-th primitive root of unity.

In  Ref. 1
the explicit matrix representations of the colored braid group
were introduced. It was shown that
the representations are equivalent
to the $R$ matrices of the nilpotent reps,
which are derived from  the universal $R$ matrix ${\cal R}$ :
$R^{a_1, a_2}_{b_1, b_2} = \pi^{p1} \otimes \pi^{p_2}
({\cal R})^{a_1a_2}_{b_1b_2}$. $^{5}$

We can show the fusion rule for the tensor product:
$V^{(p_1)} \otimes V^{(p_2)}$ =
$\sum_{p_3} N_{p_1p_2}^{p_3} V^{(p_3)}$, where
$N_{p_1,p_2}^{p_3}  =  1$
for $p_3=p_1+p_2-n$ ( $0 \le n \le N-1$ and $n \in {\bf Z}$);
$N_{p_1,p_2}^{p_3} =  0$,  otherwise.
The CGC for $V^{(p_i)}$  ($i=1,2,3$)
are given by  $^{6}$
\bea
&& C(p_1, p_2, p_3; z_1, z_2, z_3)
= \delta(z_3, z_1+z_2-n) \non \\
&& \times {\sqrt{[2p_1+2p_2-2n+1]_{\epl}}}
{\sqrt{[n]_{\epl}! [z_1]_{\epl}!  [z_2]_{\epl}!  [z_3]_{\epl}!}}
\non \\
&& \times \sum_{\nu} {\frac {(-1)^{\nu} {\epl}^{-\nu(p_1+p_2+p_3+1)}
\times {\epl}^{(n-n^2)/2 + (n-z_2)p_1 +(n+z_1)p_2} }
{[\nu]_{\epl}! [n-\nu]_{\epl}! [z_1-\nu]_{\epl}!
[z_2-n+\nu]_{\epl}!}} \non \\
&& \times {\sqrt{\frac
{[2p_1-n; z_1-\nu]_{\epl} ! [2p_1-z_1; n-\nu]_{\epl}!
 [2p_2-n; z_2+\nu -n]_{\epl} ! [2p_2-z_2; \nu]_{\epl} ! }
 {[2p_1 +2p_2 -n+1; z_1+z_2+1]_{\epl}! }}} .
\label{eq:eppp}
\eea
Here  $0 \le z_i \le N-1$, for $i$=1,2,3, and
the sum over  the integer $\nu$
in \rf{eppp} is taken
under the condition:
max $\{0, n-z_2 \}$  $\le \nu  \le$  min $\{n, z_1 \}$.
The expression of the CGC was proved
through the infinite dimensional representations. $^{6}$

\section{A Property of CGC}
We consider  the following two sums;
\bea
A^n_{p_1p_2} & = & \sum_{z_1}
\left[ C(p_1,p_2,p_3; z_1,z_2,z_1+z_2-n)\right]^2
q^{2\rho(z_1)}, \non \\
B^n_{p_1p_2} & = & \sum_{z_2} \left[ C(p_1,p_2,p_3; z_1,z_2,z_1+z_2-n)
\right]^2
q^{-2\rho(z_2)}.
\eea
For the nilpotent reps of $U_q(sl_2)$
we assume $\rho(z)=p-z+Np/2$ and $q=\epl$.
Due to the irreducibility of the representations
the sum $A^n_{p_1 p_2}$ ($B^n_{p_1p_2}$) does not depend on $z_2$
($z_1$).  We now introduce  an important  property of CGC.

\begin{prop} The CGC of the nilpotent representations satisfy
\be
A^n_{p_1 p_2} / B^{n}_{p_1 p_2} = g(p_1,p_2)=f(p_1)/f(p_2),
\mbox{ independent of } n  \quad ( p_3=p_1+p_2-n).
\label{eq:prop}
\ee
\end{prop}
(For the nilpotent reps we have
$g(p_1, p_2)=  [2p_2;N-1]_{\epl}!/[2p_1;N-1]_{\epl}!$).

{\it Proof}

We set $z_2=0$ for $A^n_{p_1p_2}$ and $z_1=0$ for $B^n_{p_1p_2}$, and
we define
$L_{p_1 p_2}^n$ and $R_{p_1 p_2}^n$ by
$A^{n}_{p_1p_2}= L_{p_1 p_2}^n [2p_3+1]/[2p_3+1+n;N]! $ and
$B^{n}_{p_1p_2} = R_{p_1 p_2}^n [2p_3+1]/[2p_3+1+n;N]! $ , respectively.
Then we can show
\be
L_{p_1p_2}^n  =  (-1)^{nN} [2p_2;N-1]_{\epl}!, \quad
R_{p_1p_2}^n  =  (-1)^{nN} [2p_1;N-1]_{\epl}!.  \label{eq:app}
\ee
The first eq. in \rf{app} can be shown by
using the following recurrence relation on $n$.
\be
L_{p_1p_2}^n =
{\frac {[2p_1+2p_2-n-N+2]}{[2p_2+1]}} {\epl}^{2p_1-n}L^{n+1}_{p_1+1/2,p_2+1/2}
- {\frac {[2p_1-n]}{[2p_2+1]}} L^{n+1}_{p_1 p_2+1/2} .
\ee
The second eq. in \rf{app}  (for $R^n_{p_1p_2}$)
is derived from that of $L_{p_1p_2}^n$ by exchanging
 $p_1$ with $p_2$, and by setting $\epl \rightarrow \epl^{-1}$.

It is easy to see that the property \rf{prop} of CGC  also holds for the
finite dimensional (spin) representations of $U_q(sl_2)$ with $q$ generic,
where  we have $g(j_1, j_2)=  [2j_1]/[2j_2]$.

We can define invariants of colored links using the proposition 3.1.
Let $T$ be a $(1,1)$- tangle.
We denote by $\hat T$ the link obtained
by closing the open strings of $T$.
We note the following proposition. $^{2}$

\begin{prop}
Let $T_1$ and $T_2$ be two $(1,1)$-tangles.
If $\hat T_1$ is isotopic to $\hat T_2$
as a link in $S^3$ by an isotopy
which carries the closing component of $\hat T_1$
to that of $\hat T_2$.
Then $T_1$ is isotopic to $T_2$ as a $(1,1)$-tangle.
\end{prop}

Let us introduce the functor $\phi(\cdot)$
for the tangle diagrams. $^{2}$
We denote by $\phi(T,\Al)^{a}_b$
the value $\phi$ for the tangle
with variables $a$ and $b$
on the closing component (or edge).
It is easy to show that
$\phi(T,\Al)^a_b = \lm \delta_{ab}$.
$^{2}$

We put $L=\hat T$
and $s$ is the color of the
closing component (or edge) of $\hat T$.
For a colored link $(L,\Al)$ and a color
$s$ of closing component (or edge),
we define $\Phi$ by $\Phi(L,s,\Al) = \lm$
where $L,T,s$ are as above  and
$\phi(T,\Al)^a_b = \lm \delta_{ab}$.
We can show that
$\Phi$ is well-defined, i.e. $\Phi(L,s,\Al)$
does not depend on a choice of $T$. $^{2}$
{}From the proposition 3.1 we have the following.
\begin{prop} $^{2}$
For a link $L$ and its color $\Al = (p_1, \cdots, p_n)$,
we have
\be
 \Phi(L,s,\Al)
([p_s; N-1]_{\epl}!)^{-1}
=\Phi(L,s',\Al)
([p_{s'}; N-1]_{\epl}!)^{-1} .
\ee
\end{prop}

\begin{df} $^{2}$
For a  colored oriented link $(L,\Al)$,
we define an isotopy invariant $\hat \Phi $ of $(F,\Al)$ by
\be
\hat \Phi(L,\Al) = \Phi (L,s,\Al)
([p_s; N-1]_{\epl}!)^{-1} .
\ee
\end{df}

Thus we have constructed the multivariable invariants from the property
\rf{prop} of
CGC. In the same way we can define the multivariable invariants of
colored ribbon graphs. $^{6}$

We now consider  CGC of $U_q(g)$, where $g$ is a simple Lie algebra.
\bea
A^{\Lambda_3}_{\Lm_1 \Lm_2}
& = & \sum_{{\vec z}_1} \left[ C(\Lm_1,\Lm_2,\Lm_n; {\vec z}_1,{\vec z}_2,
{\vec z}_1+{\vec z}_2-{\vec n})\right]^2
q^{2\rho({\vec z}_1)}, \non \\
B^{\Lm_3}_{\Lm_1 \Lm_2} & = & \sum_{{\vec z}_2}
\left[ C(\Lm_1,\Lm_2,\Lm_3; {\vec z}_1,{\vec z}_2,
{\vec z}_1+{\vec z}_2-{\vec n})\right]^2
q^{-2\rho({\vec z}_2)}.
\eea
We assume that $\rho$ is given by "half the sum of positive roots".
Finally,  we propose the following conjecture.

\begin{conj}
(1) The CGC of the nilpotent representations $\Lm_i$ of $U_q(g)$ satisfy
\be
A^{\Lm_3}_{\Lm_1 \Lm_2} / B^{\Lm_3}_{\Lm_1 \Lm_2} = g(\Lm_1,\Lm_2),
\mbox{ independent of } \Lm_3.  \label{eq:conj}
\ee
(2) With a proper normalization of the quantum trace
( $\rho({\vec z})$ $\rightarrow$ $\rho({\vec z})$ + $constant$),
we can set $g(\Lm_1,\Lm_2) = f(\Lm_1)/f(\Lm_2)$.
\par \noindent
(3) $f(\Lm)$ is equivalent to the tangle invariant for the Hopf link.
\end{conj}

If the conjecture is true,
we can construct  multivariable invariants of colored links and
colored ribbon graphs from $U_q(g)$ in the same way as  $U_q(sl_2)$.

It is easy to show that the property \rf{conj} holds for CGC of
finite dimensional representations of $U_q(g)$ with $q$ generic.

\end{document}